\documentclass[usenatbib,12pt,letterpaper,notitlepage]{article}
\usepackage[margin=1in]{geometry}
\usepackage[parfill]{parskip}    % Activate to begin paragraphs with an empty line rather than an indent
\usepackage{graphicx}
\usepackage{amssymb}
\usepackage{amsmath}
\usepackage{amsbsy}
\usepackage{epstopdf}
\usepackage{pdflscape}
\usepackage{setspace}
\usepackage{nicefrac}
\usepackage{multirow}
\usepackage{booktabs}
\usepackage{natbib}
\usepackage{enumerate}
\usepackage{makeidx}  %%% standard INDEX
\usepackage[titletoc]{appendix}
\usepackage{fancyhdr}

\usepackage{longtable}

\usepackage[latin1]{inputenc}
\usepackage{geometry}
\usepackage{graphicx}
\usepackage{epstopdf}
\usepackage{amsmath}
\usepackage{amsfonts}
\usepackage{amssymb}
\usepackage[parfill]{parskip}	% activate to begin paragraphs without an indent
\usepackage{color}
\usepackage{bm}
\usepackage{float}
\usepackage{setspace}		% 1.5 spacing
\onehalfspace
\usepackage{hyperref}
\usepackage{subfigure}

\usepackage[font=small,labelfont=bf]{caption} %activate for figure and table captions to be put in smaller font

\onehalfspace
\allowdisplaybreaks

\begin{document}

\title{\begin{large}
Rao-Blackwellization to Give Improved Estimates in\\
Multi-List Studies
\end{large}}

\author{\begin{large}Kyle Vincent\end{large}\footnote{Independent Researcher and Consultant, Ottawa, Ontario, CANADA,  \textit{email}: kyle.shane.vincent@gmail.com}}

\date{\today}

\maketitle
\pagenumbering{gobble}
\begin{abstract}
\begin{small}
\noindent Sufficient statistics are derived for the population size and capture-effect parameters of commonly used closed population mark-recapture models. Rao-Blackwellization procedures for improving on estimators that are not functions of such statistics are presented. As Rao-Blackwellization entails enumerating all sample reorderings consistent with the sufficient statistic, Markov chain Monte Carlo resampling procedures are provided to approximate the computationally intensive estimators. Simulation studies and empirical applications demonstrate that significant improvements for such estimators can be made with the strategy. Supplementary materials for this article are available online. The code will be made publicly available to facilitate further research.
\end{small}
\end{abstract}

\pagenumbering{arabic}

\clearpage

\section{Introduction}
The field of mark-recapture is a well-studied topic. Amongst the comprehensive sources of information on the subject are \cite{Williams2002} and \cite{Schwarz1999}. Mark-recapture has numerous applications to wildlife studies, and more recently for multiple systems estimation which is based on administrative lists of hidden populations like human-trafficking victims and drug-users \citep{Bird2018, Bales2015, King2013}. In this paper, sufficient statistics for the population size and parameters of closed population mark-recapture models are derived. Some mark-recapture estimators, which may or may not already be functions of these sufficient statistics, can result in extreme or unstable estimates of the population size. This is especially the case when the number of captures and/or overlap between capture occasions is nil or small, which is commonly seen in the context of multiple systems estimation \cite{Silverman2019, Chan2019, SparseMSE}. However, some commonly used mark-recapture estimators which are not functions of these sufficient statistics may result in stable estimates when Rao-Blackwellized, in particular by dampening extreme estimates so they are closer to their expectation, thereby providing a significant increase in their precision.

In mark-recapture studies the model is usually taken to be multinomial so that the population units are distributed among all possible capture histories and estimation methods are commonly carried out with respect to such a model; see \cite{Sandland1984}. In this paper, preliminary estimators are based on a variety of strategies that may or may not be directly based on the model. Such estimators do not typically depend on the unit labels. However, Rao-Blackwellization/improved estimation is based on the unit labels and the resulting improved estimators are obtained via weighing over estimates corresponding to all possible capture histories (sample reorderings) consistent with the sufficient statistic, which in turn is based on the assumed multinomial mark-recapture model.

In the event there are large sample sizes and/or a large number of sampling occasions, evaluating the improved estimators may be computationally difficult as there will likely be a prohibitively large number of reorderings to tabulate. A practical method to approximate the estimators with a Markov chain Monte Carlo (MCMC) resampling procedure is therefore provided for each mark-recapture model investigated in this paper.

The paper is organized as follows. Section 2 discusses the mark-recapture estimators used in the simulations along with the simulation study details. Section 3 outlines the estimation and variance estimation procedure for the Rao-Blackwellized estimators and their corresponding approximations with the use of a resampling algorithm. Section 4 introduces some of the nomenclature used in the paper. Results corresponding to the null, heterogeneity, behavioural, and time-effects mark-recapture models are then presented in Sections 5, 6, 7, and 8, respectively. Within each section, the notation is first introduced, followed by the sufficiency result, resampling algorithm, and results from a simulation study. The paper concludes with a discussion in Section 9. The approach outlined in this paper has been developed for the stratified mark-recapture setup, that is, where capture probabilities are assumed to be homogeneous within each stratum. The theoretical details and simulation study results can be found in the supplementary materials. Empirical applications of the Rao-Blackwell inferential strategy can also be found in the supplementary materials.

\section{Mark-Recapture Estimators and Simulation Study Details}

All simulation studies are performed in the R programming language \citep{Rprogram}.  Several classes of estimators are explored and listed as follows.
\begin{itemize}
\item The bias-adjusted Lincoln-Petersen estimator \citep{Chapman1951} and corresponding variance estimator presented in \cite{Seber1970}.
\item The maximum likelihood estimates based directly on the likelihood corresponding to the model parameters and capture histories; such estimates are detailed in \cite{Williams2002} and obtained with the \textbf{RMark} package \citep{RMark}.
\item The loglinear mark-recapture model estimates based on fitting a Poisson regression model, where the population size estimate is derived from the maximum likelihood estimates of the loglinear parameters \citep{Rivest2001}; these estimates are obtained with the \textbf{Rcapture} package \citep{Baillargeon2007, Rivest2012}.
\item Bayes estimates based on a computationally efficient semi-complete data likelihood approach, which is composed of the product of a complete likelihood which corresponds with the captured units and a marginal likelihood which corresponds with the uncaptured units, and a hybrid approach based on a data augmentation and numerical integration approximation technique applied to the semi-complete likelihood \citep{King2016, McClintock2017}; these estimates are obtained with the \textbf{multimark} package \citep{McClintock2015}.
\item The jackknife estimator detailed by \cite{Burnham1978}; this estimator is obtained with the \textbf{SPECIES} package \citep{Wang2011}.
\item The sample coverage estimates, which are based on measures of overlap and dependence between sample capture probabilities \citep{Chao1998, Chao2001, Chao2002, Chao2014}; these estimates are obtained with the \textbf{CARE1} package \citep{Hsieh2012} and \textbf{SPECIES} package \citep{Wang2011}.
\end{itemize}
All simulation studies are based on runs of three samples. A total of 2500 simulation runs are obtained for each study as these are found to sufficiently approximate the mean and variance of the preliminary estimators. The number of resamples used to approximate the mean and variance of the set of estimators corresponding to the sample reorderings consistent with the sufficient statistic for each set of three selected samples is based on Gelman-Rubin statistics and examination of visual plots of the chains \citep{Gelman1992, Gelman2004}. The number of resamples used in each simulation is noted in the corresponding simulation study subsections. Nominal 95\% confidence/probability intervals are based on the corresponding likelihood-, loglinear Poisson-, posterior-, or bootstrap estimation-based standard errors and central limit theorem. Alternative log-transformation-based intervals corresponding to the maximum likelihood and sample coverage approach are based on the method outlined in \cite{Chao1987}, and alternative equal-tailed 95\% probability intervals corresponding to the Bayes estimators are based on the posterior distributions. The alternative intervals corresponding to the improved estimators are based on averaging over the percentiles of the confidence limits or posterior distribution corresponding to the sampled reorderings, as explained in the following section.

\section{Rao-Blackwell/Improved Estimation}\label{RB_section}

Traditional/preliminary inference for population size and mark-recapture model parameters are typically based on ignoring unit labels and then basing inference directly on statistics of sample sizes, frequency counts corresponding to each possible capture history, and overlap measures between the samples; see, for example, \cite{Williams2002} for further details. Improved inference, as presented in this paper, is based on sample reorderings which in turn are based on the unit labels and full set of all possible individual capture histories; sufficiency results based on the \textit{labels-based likelihoods} are detailed in full in the following sections for the mark-recapture models explored in this paper. The purpose of using the unit labels is to aid in and simplify evaluation of the improved estimators, since combinatorial calculations corresponding to the capture histories when ignoring unit labels are not required, and hence computational issues such as machine zeros can be avoided (cf. the labels-based and traditional likelihoods in the following sections).

Suppose a set of samples are obtained for a mark-recapture study under a particular capture model. A sample reordering of the original samples is a hypothetical re-assigning of the sample units to the samples, so the number of captures over the study is equal to that originally observed and hence which gives rise to capture histories different than those originally observed; Table \ref{M0_reordering} provides an example for a sample selected under the null $M_0$ model (that is, where capture probabilities are equal and constant for all units across sampling occasions).
\begin{table}[H]
\centering
\caption{Left: Original sample selected under $M_0$ mark-recapture model. Right: A sample reordering.}
\begin{tabular}{l||lll||llllll}\hline
Unit                        &Sample 1 &Sample 2   &Sample 3       &Sample 1 &Sample 2   &Sample 3 \\\hline
$A$                         &1        &1          &0              &1        &1          &1      \\
$B$                         &0        &1          &1              &0        &1          &0      \\
$C$                         &1        &1          &0              &1        &0          &1      \\
\hline
\end{tabular}
\label{M0_reordering}
\end{table}

\vspace*{-\baselineskip}
Define $d_R$ to be a sufficient statistic for a population quantity or model parameter $\gamma$, such as the population size or a capture probability corresponding to the mark-recapture model,. For the setup considered in this paper the sample reorderings that are mapped to the same $d_R$ as that which corresponds with the original sample orderings are consistent with the sufficient statistic and therefore are those which contribute to improved estimation. Let $\mathcal{R}$ be the set of all such sample reorderings. Index the sample reorderings as 1,...,$|\mathcal{R}|$. Define $d_0^{[r]}$ to be the hypothetical samples corresponding to reordering $r$; for example, for some $r$ the set $d_0^{[r]}$ would be the samples displayed in the right side of the example presented in Table \ref{M0_reordering}. Define $\hat{\gamma}_0^{[r]}$ to be the estimate of $\gamma$ based on the set of samples in $d_0^{[r]}$. Finally, define $p(d_0^{[r]})$ to be the probability of observing the reordered samples under the assumed capture model based on the observed unit labels and in the order the samples are selected. It can be seen that with the sufficiency results based on the labels-based likelihoods given in this paper, under any of the models $p(d_0^{[r]})$ is uniform amongst all reorderings which are consistent with the corresponding sufficient statistic; details and examples are provided in the following sections. Hence, the improved estimate is the arithmetic mean of the estimates that correspond to sample reorderings that could have been observed under the assumed mark-recapture model and which are consistent with the sufficient statistic; the improved estimate of $\gamma$ is
\begin{align}
\hat{\gamma}_{RB}&=E[\hat{\gamma}_0|d_r]
=\sum\limits_{i\epsilon \mathcal{R}}\bigg(\hat{\gamma}_0^{[r]}p(d_0^{[r]}|d_r)\bigg)
=\frac{\sum\limits_{i\epsilon \mathcal{R}}\bigg(\hat{\gamma}_0^{[r]}p(d_0^{[r]})\bigg)}{\sum\limits_{i\epsilon \mathcal{R}}p(d_0^{[r]})}
=\frac{\sum\limits_{i\epsilon \mathcal{R}}\hat{\gamma}_0^{[r]}}{|\mathcal{R}|}.
\label{Mh_RB_expression}
\end{align}

To estimate the variance of the improved estimator, the decomposition of variances gives
\begin{align}
\text{var}(\hat{\gamma}_{RB})=\text{var}(\hat{\gamma}_0)-E[\text{var}(\hat{\gamma}_0|d_r)].
\label{varest1}
\end{align}
If $\hat{\text{var}}(\hat{\gamma}_0)$ is an estimator of $\text{var}(\hat{\gamma}_0)$ then an estimator of $\text{var}(\hat{\gamma}_{RB})$ is
\begin{align}
\hat{\text{var}}(\hat{\gamma}_{RB})=E[\hat{\text{var}}(\hat{\gamma}_0)|d_r]-\text{var}(\hat{\gamma}_0|d_r).
\label{varest2}
\end{align}
This estimator is the difference of the expectation of the estimated variance of the preliminary estimator over all consistent reorderings and the variance of the preliminary estimator over all consistent reorderings. Although these estimates are unbiased, they can result in negative estimates of the variance. For such a case, a conservative approach is to set the estimate of $\text{var}(\hat{\gamma}_{RB})$ equal to $E[\hat{\text{var}}(\hat{\gamma}_0)|d_r]$. This approach is utilized in the simulation studies.

With the Markov chain Monte Carlo (MCMC) resampling procedures outlined in this paper, approximations to the Rao-Blackwellized estimators can be obtained as follows. Suppose the number of resamples/length of the MCMC chain is $M$. Define $q(d_0^{[r]})$ to be the probability of selecting sample reordering $r$ under the proposal distribution. Suppose that at step $m$ of the chain the most recently accepted sample reordering drawn from the proposal distribution is $d_0^{[r^*]}$ for some $r^*$. Draw a candidate sample reordering, $d_0^{[r]}$ say, from the proposal distribution. With probability $\alpha=\text{min}\bigg\{\frac{p(d_0^{[r]})}{p(d_0^{[r^*]})}\frac{q(d_0^{[r^*]})}{q(d_0^{[r]})},1\bigg\}=\text{min}\bigg\{\frac{q(d_0^{[r^*]})}{q(d_0^{[r]})},1\bigg\}$ accept the candidate sample reordering $d_0^{[r]}$ for step $m$, and with probability $1-\alpha$ reject the sample reordering and retain $d_0^{[r^*]}$ for step $m$.

Let $\hat{\gamma}^{(m)}_0$ be the preliminary estimate of $\gamma$ obtained with the most recently accepted sample reordering selected at step $m$. An enumerative estimate of $\hat{\gamma}_{RB}$ is then
\begin{align}
\tilde{\gamma}_{RB}=\frac{\sum\limits_{m=1}^M\hat{\gamma}^{(m)}_0}{M}.
\end{align}
\noindent Similarly, let $\hat{\text{var}}(\hat{\gamma}^{(m)}_0)$ be the estimate of the variance of $\hat{\gamma}_0$ obtained with the most recently accepted sample reordering selected at step $m$. An enumerative estimate of $\hat{\text{var}}(\hat{\gamma}_{RB})$ is then
\begin{align}
\tilde{\text{var}}(\hat{\gamma}_{RB})&
=\tilde{E}[\hat{\text{var}}(\hat{\gamma}_0)\mid d_r]-\tilde{\text{var}}(\hat{\gamma}_0\mid d_r)\notag \\
&=\frac{1}{M}\sum\limits_{m=1}^M\hat{\text{var}}(\hat{\gamma}_{0}^{(m)})-\frac{1}{M}\sum\limits_{m=1}^M (\hat{\gamma}_{0}^{(m)}-\tilde{\gamma}_{RB})^2.
\end{align}

\section{Nomenclature}

The following notation is used in this paper. Define $N$ to be the population size and $K$ the number of sampling occasions for the mark-recapture study. Define $U=\{1,2,...,N\}$ to be the set of population unit labels, $s_k$ to be the set of units captured on sampling occasion $k$, $n_k=|s_k|$, $s=\cup_{k=1}^K s_k$, and $n=|s|$. Define $p_{ik}$ to be the capture probability of unit $i$ on sampling occasion $k$, $C_{i,k}=1$ if unit $i$ is captured on sampling occasion $k$ and zero otherwise, and $C_i=\sum\limits_{k=1}^KC_{i,k}$ the total number of times unit $i$ is captured over the study. For the purposes of deriving the sufficiency results based on the labels-based likelihoods, and without loss of generality, the unit labels of $s$ are taken to be $1,2,...,n$. Define $I[s\subseteq U]=1$ if $s\subseteq U$ and 0 otherwise, where in this setup $U$ is considered to be a function of $N$. Finally, define $\{x_{\omega}\}$ to be the number of individuals exhibiting each capture history $\omega$. For example, if $K=3$ and $\omega=(1,0,1)$ then $x_{\omega}$ is the number of individuals captured on occasions 1 and 3, but not on occasion 2.

\section{Null Model}

Under the $M_0$ model, $p_{ik}=p$ for all $i=1,2,...,N$ and $k=1,2,...,K$. Define $C$ to be the total number of captures in the study.

\subsection{Sufficiency Results}

When considering unit labels, define the original data to be $d_0=\{s_k:k=1,...,K\}$ and the reduced data to be $d_R=\{s,C\}$.

\textbf{Theorem:} The reduced data $D_R$ is sufficient for $(N,p)$.

\textbf{Proof:} For any $(d_0, p)$, the labels-based likelihood is
\begin{align}
\notag
&P_{N}(D_0=d_0)=P(s_1,s_2,...,s_K)I[s\subseteq \{1,2,...,N\}]\\\notag
&=\prod\limits_{k=1}^K\bigg[p^{\sum\limits_{i \epsilon s}C_{i,k}}
(1-p)^{\sum\limits_{i\epsilon s}{(1-C_{i,k})}}  (1-p)^{(N-n)}   \bigg]I[s\subseteq \{1,2,...,N\}]\\\notag
&=p^{\sum\limits_{k=1}^K\sum\limits_{i \epsilon s}C_{i,k}}
(1-p)^{\sum\limits_{k=1}^K\sum\limits_{i\epsilon s} (1-C_{i,k})}  (1-p)^{(KN-Kn)} I[s\subseteq \{1,2,...,N\}]\\\notag
&=\bigg[p^{C}(1-p)^{(KN-C)}  \bigg]I[s\subseteq \{1,2,...,N\}]\\\notag
&=g(N,p,d_R)h(d_0)\\
\label{likelihood_M0}
\end{align}
where $h(d_0)=1$. Therefore, by the Neyman-Factorization Theorem $D_R$ is sufficient for $(N,p)$. $\Box$

As presented in \cite{Williams2002}, when ignoring unit labels the likelihood for $(N,p)$ based on the capture histories is
\begin{align}
\notag
P(\{x_{\omega}\}|N,p)&=\frac{N!}{\bigg[\prod\limits_{\omega}x_{\omega}!\bigg](N-n)!}p^C(1-p)^{(KN-C)}\\\notag
&=\frac{N!}{(N-n)!}p^C(1-p)^{(KN-C)}\frac{1}{\prod\limits_{\omega}x_{\omega}!}\\\notag
&=g(N,p,T(\{x_{\omega}\}))h(\{x_{\omega}\})\\
\label{traditional_likelihood_M0}
\end{align}
where $h(\{x_{\omega}\})=\frac{1}{\prod\limits_{\omega}x_{\omega}!}$. Therefore, by the Neyman Factorization theorem, $T(\{x_{\omega}\})=(n,C)$ is the analogous sufficient for $(N,p)$.$\Box$

Table \ref{M0_reordering} in Section \ref{RB_section} depicts a sample selected under the $M_0$ model (where A corresponds with unit 1, B with unit 2, and C with unit 3), where the probability of observing the original data is $p^6(1-p)^{(3N-6)}$, and a sample reordering that is consistent with the reduced data.

\subsection{Resampling Procedure}

A Metropolis-Hastings MCMC chain \citep{Hastings1970} is used to approximate the improved estimators with the following proposal distribution.

Step 1: For each of the $n$ individuals in $s$, assign them a capture to a randomly chosen sample and tabulate the capture history matrix.

Step 2: Distribute the remaining $\sum\limits_{k=1}^Kn_k - n$ captures at random to the zero entries in the capture history matrix.

In the first step and for each captured unit $i$, there are $C_i$ possible ways of first assigning this unit to one of the $C_i$ samples in which they are captured in the sample reordering. Therefore, the total possible number of ways of selecting the proposed sample reordering is $\prod\limits_{i\ \epsilon\ s} C_i$. For the accept/reject portion of the chain, the sample reordering is accepted with probability $\text{min}\bigg\{\frac{q(d_0^{[r^*]})}{q(d_0^{[r]})},1\bigg\}=\text{min}\bigg\{\frac{\prod\limits_{i\ \epsilon\ s} C_i^*}{\prod\limits_{i\ \epsilon\ s} C_i}, 1\bigg\}$ where $d_0^{[r^*]}$ and $d_0^{[r^]}$ respectively correspond to the most recently accepted sample reordering and proposed sample reordering.

\subsection{Simulation Study}

The population size is set to $N=500$ and capture probability to $p=0.15$. The following estimators, which are already functions of the sufficient statistic, are used in the simulation study: the maximum likelihood estimator based directly on the capture histories, maximum-likelihood estimator based on a Poisson log-likelihood, and Bayes estimator, all three of which are based on the assumption of no capture effects. The following estimators, which are not functions of the sufficient statistic and can therefore benefit from the Rao-Blackwellization procedure detailed in this section, are also used in the simulation study: the bias-adjusted Lincoln-Petersen estimator, the jackknife estimator, and the Chao-Tsay sample coverage approach estimator. Table \ref{Sim_M0} presents the results from the simulation study. Improved estimates are based on 2500 resamples. The acceptance rate of the MCMC chain is 83.14\%.

\begin{longtable}{l*{10}{r}r}
\caption{Mean, variance, mean-squared error, coverage rates, average length of confidence intervals, coverage rates of alternative confidence intervals, average length of alternative confidence intervals, and percentage of variance estimates which are negative, for preliminary and improved estimators. Top: Estimators which are already functions of the sufficient statistic. Middle: Preliminary estimators. Bottom: Improved estimators.}
\endfirsthead
\multicolumn{10}{l}
{{Table \ref{Sim_M0} continued from previous page}} \\
  \hline
Estimator           &Mean           &Var.           &MSE                &CR      &Length   &Alt. CR       &Alt. Length         &Neg. Est.             \\\hline
\endhead
Estimator           &Mean           &Var.           &MSE                &CR      &Length   &Alt. CR       &Alt. Length         &Neg. Est.            \\\hline
MLE                 &509            &5,972          &6,052              &0.943   &293      &0.951         &299                 &NA              \\
MLE LL              &501            &5,390          &5,391              &0.933   &283      &NA            &NA                  &NA              \\
Bayes               &511            &5,735          &5,856              &0.950   &296      &0.950         &295                 &NA              \\\hline

LP                  &501            &19,747         &19,748             &0.896   &489      &NA            &NA                  &NA              \\
JK                  &464            &3,696          &5,016              &0.751   &113      &NA            &NA                  &NA              \\
SC CT               &515            &9,895          &10,124             &0.950   &399      &0.963         &414                 &NA              \\\hline

LP RB               &500            &5,297          &5,297              &0.931   &283      &NA            &NA                  &0\%                 \\
JK RB               &464            &865            &2,152              &0.708   &109      &NA            &NA                  &90.56\%                   \\
SC CT RB            &514            &6,136          &6,338              &0.964   &343      &NA            &NA                  &0\%               \\\hline
\label{Sim_M0}
\end{longtable}
\vspace*{-\baselineskip}

\section{Heterogeneity Model}

Under the $M_h$ model, $p_{ik}=p_i$ for all $i=1,2,...,N$ and $k=1,2,...,K$. Define $\underline{C}=(C_1,C_2,...,C_n)$.

\subsection{Sufficiency Results}

When ignoring unit labels, \cite{Williams2002} present the likelihood for $(N,\underline{p})$ via conceptualizing the capture probabilities as a random sample from a probability distribution. For the setup presented in this paper, the model is parameterized with $N$ capture probabilities $p_1, ..., p_N$ to facilitate a Rao-Blackwellization improvement procedure.

When considering unit labels, define the original data to be $d_0=\{s_k:k=1,...,K\}$ and the reduced data to be $d_R=\{s,\underline{C}\}$.

\textbf{Theorem:} The reduced data $D_R$ is sufficient for $(N, \underline{p})$, where $\underline{p}=(p_1, p_2, ..., p_N)$.

\textbf{Proof:} For any $(d_0,\underline{p})$, the labels-based likelihood is
\begin{align}
\notag
&P_{N}(D_0=d_0)=P(s_1,s_2,...,s_K)I[s\subseteq U]\\\notag
&=\prod\limits_{k=1}^K\bigg[\prod\limits_{i \epsilon s}\bigg[p_i^{C_{i,k}}(1-p_i)^{(1-C_{i,k})}\bigg]\bigg] \prod\limits_{k=1}^K\bigg[\prod\limits_{i=n+1}^N(1-p_i)\bigg]I[s\subseteq U]\\\notag
&=\prod\limits_{i \epsilon s}\bigg[p_i^{C_{i}}(1-p_i)^{(K-C_{i})}\bigg] \bigg[\prod\limits_{i=n+1}^N(1-p_i)^K\bigg]I[s\subseteq U]\\\notag
&=g(N,\underline{p},d_R)h(d_0)\\
\label{sufficiency_Mh}
\end{align}
where $h(d_0)=1$. Therefore, by the Neyman-Factorization Theorem $D_R$ is sufficient for $(N,\underline{p})$. $\Box$

Table \ref{Mh_reordering} depicts a sample selected under the $M_h$ model (where A corresponds with unit 1, B with unit 2, and C with unit 3), where the probability of observing the original data is $p_1^2(1-p_1) \times p_2^2(1-p_2) \times p_3^2(1-p_3) \times \prod\limits_{i=4}^N(1-p_i)^3$, and a sample reordering that is consistent with the reduced data.
\begin{table}[H]
\centering
\caption{Left: Original sample selected under $M_h$ mark-recapture model. Right: Sample reordering consistent with the reduced data.}
\begin{tabular}{l||lll||llllll}\hline
Unit                        &Sample 1 &Sample 2   &Sample 3       &Sample 1 &Sample 2   &Sample 3 \\\hline
$A$                         &1        &1          &0              &1        &0          &1      \\
$B$                         &0        &1          &1              &1        &1          &0      \\
$C$                         &1        &1          &0              &0        &1          &1      \\
\hline
\end{tabular}
\label{Mh_reordering}
\end{table}

\subsection{Resampling Procedure}
Sample reorderings are selected completely at random with the following algorithm to approximate the improved estimators. Essentially, for each captured unit their corresponding capture history is permuted amongst the $K$ samples to give rise to a sampled reordering. Under an MCMC setup, all sample reorderings have equal probability of being selected under the proposal distribution. Therefore, an accept/reject step is avoided.

\subsection{Simulation Study}
The population size is set to $N=500$ and capture probabilities are generated according to a Beta (3,10) distribution. The following estimators, which are already functions of the sufficient statistic, are used in the simulation study:  Chao's lower bound and Poisson estimator based on a Poisson log-likelihood, both of which are based on the assumption of heterogeneity effects, the jackknife estimator, and the Chao-Bunge sample coverage approach estimator. The following estimators, which are not functions of the sufficient statistic and can therefore benefit from Rao-Blackwellization, are also used in the simulation study: the Bayes estimator, based on the assumption of heterogeneity effects, and the Chao-Tsay sample coverage approach estimator. Table \ref{Sim_Mh} presents results from the simulation study. Improved estimates are based on 1000 resamples.

\begin{longtable}{l*{10}{r}r}
\caption{Mean, variance, mean-squared error, coverage rates, average length of confidence intervals, coverage rates of alternative confidence intervals, average length of alternative confidence intervals, and percentage of variance estimates which are negative, for preliminary and improved estimators. Top: Estimators which are already functions of the sufficient statistic. Middle: Preliminary estimators. Bottom: Improved estimators.}
\endfirsthead
\multicolumn{10}{l}
{{Table \ref{Sim_Mh} continued from previous page}} \\
  \hline
Estimator       &Mean           &Var.           &MSE                &CR      &Length   &Alt. CR       &Alt. Length         &Neg. Est.             \\\hline
\endhead
Estimator       &Mean           &Var.           &MSE                &CR      &Length   &Alt. CR       &Alt. Length         &Neg. Est.            \\\hline
Chao            &428            &1,275          &6,501              &0.421   &134      &NA            &NA                  &NA                  \\
Poisson         &455            &5,112          &7,096              &0.779   &266      &NA            &NA                  &NA                  \\
JK              &560            &1,018          &4,668              &0.553   &130      &NA            &NA                  &NA                  \\
SC CB           &444            &2,641          &5,785              &0.762   &190      &NA            &NA                  &NA                  \\\hline

Bayes           &549            &11,260         &13,617             &0.962   &474      &0.953         &466                 &NA                  \\
SC CT           &440            &1,970          &5,561              &0.654   &172      &0.775         &176                 &NA                  \\\hline

Bayes RB        &547            &8,703          &10,934             &0.985   &463      &0.987         &465                 &0\%                    \\
SC CT RB        &440            &1,959          &5,554              &0.643   &171      &0.782         &176                 &0\%                    \\\hline
\label{Sim_Mh}
\end{longtable}
\vspace*{-\baselineskip}

\section{Behavioural Model}

Under the $M_b$ model, $p_{ik}=p$ until first capture and $p_{ik}=\phi p$ for any recapture for all $i=1,...,N$ and $k=1,...,K$, where $\phi$ is the behavioural effect parameter. Define $m_{i,k}=1$ if unit $i$ is captured at least once before sampling occasion $k$ and 0 otherwise (by definition, $m_{i,1}=0$), and $F_k$ and $R_k$ respectively to be the number of first time captures and recaptures on sampling occasion $k$. Define $F=\sum\limits_{k=1}^K\sum\limits_{k'<k}F_{k'}$, the sum over all sampling occasions of the number of previously captured individuals, and $R=\sum\limits_{k=1}^KR_k$, the total number of recaptures.

\subsection{Sufficiency Results}

When considering unit labels, define the original data to be $d_0=\{s_k:k=1,...,K\}$ and the reduced data to be $d_R=\{s,F,R\}$.

\textbf{Theorem:} The reduced data $D_R$ is sufficient for $(N,\phi,p)$.

\textbf{Proof:} For any $(d_0,p,\phi)$, the labels-based likelihood is
\begin{align}
\notag
&P_{N}(D_0=d_0)=P(s_1,...,s_K)I[s\subseteq U]\\\notag
&=\prod\limits_{k=1}^K\bigg[\prod\limits_{i \epsilon s} \bigg( (\phi^{m_{i,k}}p)^{C_{i,k}}(1-\phi^{m_{i,k}}p)^{(1-C_{i,k})}\bigg)(1-p)^{(N-n)}\bigg] I[s\subseteq U]\\\notag
&=\prod\limits_{k=1}^K\bigg[\phi^{\sum\limits_{i \epsilon s}(m_{i,k}C_{i,k})}p^{\sum\limits_{i \epsilon s}C_{i,k}} \prod\limits_{i \epsilon s}\bigg[(1-\phi^{m_{i,k}}p)^{(1-C_{i,k})}\bigg](1-p)^{(N-n)}\bigg]I[s\subseteq U]\\\notag
&=\prod\limits_{k=1}^K\bigg[\phi^{R_k}p^{F_k}p^{R_k}\bigg] \prod\limits_{k=1}^K\bigg[(1-p)^{\sum\limits_{k' > k}F_{k'}}(1-p)^{N-n}(1-\phi p)^{(\sum\limits_{k' < k}F_{k'}-R_k)}\bigg]I[s\subseteq U]\\\notag
&=(\phi p)^{R}p^{n}(1-\phi p)^{(F-R)}(1-p)^{(KN-n-F)}I[s\subseteq U]\\\notag
&=g(N,p,\phi,d_R)h(d_0)\\
\label{sufficiency_Mb}
\end{align}
where $h(d_0)=1$. Therefore, by the Neyman-Factorization Theorem $D_R$ is sufficient for $(N,\phi,p)$. $\Box$

As presented in \cite{Williams2002}, when ignoring the unit labels the likelihood for $(N,\phi,p)$ based on the capture histories is
\begin{align}
\notag
&P(\{x_{\omega}\}|N,p,\phi)=\frac{N!}{[\prod\limits_{\omega}x_{\omega}!](N-n)!}(\phi p)^{R}p^{n} (1-p)^{(KN-n-F)}(1-\phi p)^{(F-R)}\\\notag
&=\frac{N!}{(N-n)!}(\phi p)^{R}p^{n} (1-p)^{(KN-n-F)}(1-\phi p)^{(F-R)}\frac{1}{\prod\limits_{\omega}x_{\omega}!}\\\notag
&=g(N,p,\phi,T(\{x_{\omega}\}))h(\{x_{\omega}\})\\
\label{traditional_likelihood_Mb}
\end{align}
where $h(\{x_{\omega}\})=\frac{1}{\prod\limits_{\omega}x_{\omega}!}$. It can be seen that by the Neyman Factorization theorem, the analogous sufficient statistic for $(N,\phi,p)$ is $T(\{x_{\omega}\})=\{n,F,R\}$.

Rao-Blackwellization based on the sufficient statistic $\{s,F,R\}$ poses challenges, primarily due to the cumulative sum of the first time captures that defines the statistic $F$. Therefore, for the purposes of Rao-Blackwellization, a finer and easier sufficient statistic to work with for evaluating/proposing consistent sample reorderings is used to obtain improved estimates, namely $\{s,F_k,R_k:k=1,...,K\}$.

Table \ref{Mb_reordering} depicts a sample selected under the $M_b$ capture model (where A corresponds with unit 1, B with unit 2, and C with unit 3), where the probability of observing the original data is $\phi^3p^6(1-\phi p)^2(1-p)^{(3N-8)}$, and a sample reordering that is consistent with the reduced data.
\begin{table}[H]
\centering
\caption{Left: Original sample selected under $M_b$ mark-recapture model. Right: Sample reordering consistent with the reduced data.}
\begin{tabular}{l||lll||llllll}\hline
Unit                        &Sample 1 &Sample 2   &Sample 3       &Sample 1 &Sample 2   &Sample 3 \\\hline
$A$                         &1        &1          &0              &1        &1          &1      \\
$B$                         &0        &1          &1              &0        &1          &0      \\
$C$                         &1        &1          &0              &1        &1          &0     \\
\hline
\end{tabular}
\label{Mb_reordering}
\end{table}

\subsection{Resampling Procedure}
Sample reorderings are selected completely at random with the following algorithm to approximate the improved estimators.

Step $1$: For $k=1$, sample $F_1$ units at random from $s$ and assign them to sample 1. Denote the units selected for sample 1 at this step as $s_1'$.

For each $k=2,...,K$, sample $F_k$ units at random from $s\setminus \cup_{j=1}^{k-1} s_j'$ and assign them to sample $k$.  Denote the units selected for sample $k$ at this step as $s_k'$.

Step $2$: For each $k=2,...,K$, sample $R_k$ units at random from $\cup_{j=1}^{k-1} s_j \setminus s_k'$ and assign them to sample $k$.

Under an MCMC setup, all sample reorderings have equal probability of being selected under the proposal distribution. Therefore, an accept/reject step is avoided.

\subsection{Simulation Study}
The population size is set to $N=500$, capture probability to $p=0.3$, and behavioural effect to $\phi=1.15$. The following estimators, which are already functions of the sufficient statistic, are used in the simulation study: the maximum likelihood estimator based directly on the capture histories, maximum-likelihood estimator based on a Poisson log-likelihood, and Bayes estimator, all three of which are based on the assumption of a behavioural effect. The following estimators, which are not functions of the sufficient statistic and can therefore benefit from Rao-Blackwellization, are also used in the simulation study: the Chao-Bunge sample coverage approach estimator and the Chao-Tsay sample coverage approach estimator. Table \ref{Sim_Mb} presents results from the simulation study. Improved estimates are based on 1000 resamples.

\begin{longtable}{l*{10}{r}r}
\caption{Mean, variance, mean-squared error, coverage rates, average length of confidence intervals, coverage rates of alternative confidence intervals, average length of alternative confidence intervals, and percentage of variance estimates which are negative, for preliminary and improved estimators. Top: Estimators which are already functions of the sufficient statistic. Middle: Preliminary estimators. Bottom: Improved estimators.}
\endfirsthead
\multicolumn{10}{l}
{{Table \ref{Sim_Mb} continued from previous page}} \\
  \hline
Estimator           &Mean           &Var.           &MSE                &CR      &Length   &Alt. CR       &Alt. Length         &Neg. Est.             \\\hline
\endhead
Estimator           &Mean           &Var.           &MSE                &CR      &Length   &Alt. CR       &Alt. Length         &Neg. Est.            \\\hline
MLE                 &508            &4,430          &4,495              &0.914   &247      &0.950         &258                 &NA        \\
MLE LL              &502            &4,035          &4,040              &0.902   &236      &NA            &NA                  &NA        \\
Bayes               &527            &5,159          &5,909              &0.961   &295      &0.950         &289                 &NA        \\\hline

SC CB               &468            &987            &2,001              &0.810   &119      &NA            &NA                  &NA        \\
SC CT               &467            &1,108          &2,165              &0.767   &129      &0.874         &132                 &NA        \\\hline

SC CB RB            &469            &607            &1,596              &0.698   &95      &NA             &NA                  &0\%          \\
SC CT RB            &468            &597            &1,627              &0.694   &96      &0.939          &132                 &0\%          \\\hline
\label{Sim_Mb}
\end{longtable}
\vspace*{-\baselineskip}

\section{Time-Effects Model}

Under the $M_t$ model, $p_{ik}=pe_k$ for all $i=1,...,N$ and $k=1,...,K$, where $\underline{e}=(e_1,e_2,...,e_K)$ are the time-effect parameters.

\subsection{Sufficiency Results}

When considering unit labels, define the original data to be $d_0=\{s_k:k=1,...,K\}$ and the reduced data to be $d_R=\{s,n_k:k=1,...,K\}$.

\textbf{Theorem:} The reduced data $D_R$ is sufficient for $(N, p, \underline{e})$.

\textbf{Proof:} For any $(d_0, p, \underline{e})$, the labels-based likelihood is
\begin{align}
\notag
&P_{N}(D_0=d_0)=P(s_1,...,s_K)I[s\subseteq \{1,...,N\}]\\\notag
&=\prod\limits_{k=1}^K\bigg[\prod\limits_{i \epsilon s}\bigg((pe_k)^{C_{i,k}}(1-pe_k)^{(1-C_{i,k})}\bigg)(1-pe_k)^{(N-n)}\bigg]I[s\subseteq \{1,...,N\}]\\\notag
&=\prod\limits_{k=1}^K\bigg[(pe_k)^{n_k}(1-pe_k)^{(n-n_k)}(1-pe_k)^{(N-n)}\bigg]I[s\subseteq \{1,...,N\}]\\\notag
&=\prod\limits_{k=1}^K\bigg[(pe_k)^{n_k}(1-pe_k)^{(N-n_k)}\bigg]I[s\subseteq \{1,...,N\}]\\\notag
&=g(N,p,\underline{e},d_R)h(d_0)\\
\label{sufficiency_Mt}
\end{align}
where $h(d_0)=1$. Therefore, by the Neyman-Factorization Theorem $D_R$ is sufficient for $(N,p,\underline{e})$. $\Box$

As presented in \cite{Williams2002}, the labels-based likelihood for $(N,p,\underline{e})$ based on the capture histories is
\begin{align}
\notag
&P(\{x_{\omega}\}|N,p,\underline{e})=\frac{N!}{[\prod\limits_{\omega}x_{\omega}!](N-n)!}\prod\limits_{k=1}^K\bigg[(pe_k)^{n_k}(1-pe_k)^{(N-n_k)}\bigg]\\\notag
&=\frac{N!}{(N-n)!}\prod\limits_{k=1}^K\bigg[(pe_k)^{n_k}(1-pe_k)^{(N-n_k)}\bigg]\frac{1}{\prod\limits_{\omega}x_{\omega}!}\\\notag
&=g(N,p,\underline{e},T(\{x_{\omega}\}))h(\{x_{\omega}\})\\
\label{sufficiency_Mt}
\end{align}
where $h(\{x_{\omega}\})=\frac{1}{\prod\limits_{\omega}x_{\omega}!}$. Therefore, by the Neyman Factorization theorem, $T(\{x_{\omega}\})=(n_1,...,n_K)$ is the analogous sufficient statistic for $(N,p,\underline{e})$. $\Box$

Table \ref{Mt_reordering} depicts a sample selected under the $M_t$ model (where A corresponds with unit 1, B with unit 2, and C with unit 3), where the probability of observing the original data is
$(pe_1)^2(1-pe_1)^{(N-2)} \times (pe_2)^2 (1-pe_2)^{(N-2)} \times (pe_3)(1-pe_3)^{(N-1)}$, and a sample reordering that is consistent with the reduced data.
\begin{table}[H]
\centering
\caption{Left: Original sample selected under $M_t$ mark-recapture model. Right: Sample reordering consistent with the reduced data.}
\begin{tabular}{l||lll||llllll}\hline
Unit                        &Sample 1 &Sample 2   &Sample 3       &Sample 1 &Sample 2   &Sample 3 \\\hline
$A$                         &1        &0          &1              &1        &1          &1      \\
$B$                         &1        &0          &1              &0        &0          &1      \\
$C$                         &0        &1          &0              &1        &0          &0     \\
\hline
\end{tabular}
\label{Mt_reordering}
\end{table}

\subsection{Resampling Procedure}

A Metropolis-Hastings MCMC chain \citep{Hastings1970} is used to approximate the improved estimators with the following symmetric proposal distribution.

Step 1: Sample an entry in the capture matrix which corresponds with a capture (i.e. a one) and assign it a miss (i.e. a zero).

Step 2: For those units with a corresponding miss for this column (sample), choose one at random and assign it a capture.

Step 3: Check that all units are captured at least once in the study.

For the accept/reject portion of the chain, the sample reordering is only rejected if any units are missing from the final sample.

\subsection{Simulation Study}

The population size is set to $N=500$ and capture probabilities to $pe_1=0.30, pe_2=0.20$, and $pe_3=0.10$. The following estimators, which are already functions of the sufficient statistic, are used in the simulation study: the maximum likelihood estimator based directly on the capture histories, maximum-likelihood estimator based on a Poisson log-likelihood, and Bayes estimator, all three of which are based on the assumption of time effects. The following estimators, which are not functions of the sufficient statistic and can therefore benefit from the Rao-Blackwellization procedure detailed in this section, are also used in the simulation study: the bias-adjusted Lincoln-Petersen estimator, the jackknife estimator, the Chao-Bunge sample coverage approach estimator, and the Chao-Tsay sample coverage approach estimator. Table \ref{Sim_Mt} presents the results from the simulation study. Improved estimates are based on 15,000 resamples.  The acceptance rate of the MCMC chain is 33.96\%.

\begin{longtable}{l*{10}{r}r}
\caption{Mean, variance, mean-squared error, coverage rates, average length of confidence intervals, coverage rates of alternative confidence intervals, average length of alternative confidence intervals, and percentage of variance estimates which are negative, for preliminary and improved estimators. Top: Estimators which are already functions of the sufficient statistic. Middle: Preliminary estimators. Bottom: Improved estimators.}
\endfirsthead
\multicolumn{10}{l}
{{Table \ref{Sim_Mt} continued from previous page}} \\
  \hline
Estimator           &Mean           &Var.           &MSE                &CR      &Length   &Alt. CR       &Alt. Length         &Neg. Est.             \\\hline
\endhead
Estimator           &Mean           &Var.           &MSE                &CR      &Length   &Alt. CR       &Alt. Length         &Neg. Est.            \\\hline
MLE                 &504            &2,676          &2,689              &0.959   &205      &0.951         &208                 &NA             \\
MLE LL              &504            &2,591          &2,603              &0.960   &203      &NA            &NA                  &NA             \\
Bayes               &508            &2,712          &2,769              &0.966   &209      &0.950         &208                 &NA             \\\hline

LP                  &502            &5,047          &5,050              &0.940   &268      &NA            &NA                  &NA             \\
JK                  &592            &2,026          &10,490             &0.163   &133      &NA            &NA                  &NA             \\
SC CB               &541            &15,928         &17,641             &0.973   &383      &NA            &NA                  &NA            \\
SC CT               &508            &5,418          &5,482              &0.954   &290      &0.956         &299                 &NA             \\\hline

LP RB               &501            &2,618          &2,619              &0.950   &201      &NA            &NA                  &0\%              \\
JK RB               &592            &884            &9,291              &0.117   &109      &NA            &NA                  &0\%                  \\
SC CB RB            &542            &3,676          &5,472              &0.991   &331      &NA            &NA                  &41.10\%               \\
SC CT RB            &510            &2,760          &2,856              &0.978   &228      &0.994         &300                 &0\%              \\\hline
\label{Sim_Mt}
\end{longtable}
\vspace*{-\baselineskip}

\section{Discussion}

In this paper the mathematical details for Rao-Blackwellizing estimates of population size and mark-recapture model parameters are presented for several closed mark-recapture models. The simulation studies demonstrate that the improved estimators serve as competitive alternatives to estimators which are already functions of the sufficient statistics and hence which are naturally an average of estimates corresponding to all the consistent sample reorderings. Further, the empirical study results presented in the supplementary materials demonstrate that, under each mark-recapture model, in most cases significant reductions in the standard errors can be expected for the improved counterparts of such estimators.

Approximations to the improved estimates is aided with the likelihood based on the unit labels. If one were to use the likelihood that is not based on the unit labels, then the combinatorial calculations corresponding with the capture histories of sample reorderings would need to be calculated in order to evaluate the improved estimators. Further, the probability of observing sample reorderings will not necessarily be homogeneous, and a more complicated resampling algorithm and Monte Carlo Markov chain procedure may be needed to approximate the improved estimators. These may present computational burdens, which therefore motivates the use of the sufficiency result based on the likelihood which makes use of the unit labels.

As with all statistical modeling, it is important to base model choice on goodness-of-fit statistics and visual illustrations of the residuals. For the mark-recapture models explored in this paper there is a wealth of such model selection aids in \cite{Baillargeon2007} and \cite{RMark}, and it is suggested to base improved inference on the model that best fits the data. For cases where the chosen model may be incorrect, future work on quantifying the sensitivity of the Rao-Blackwellized estimators to departures from the chosen model would be invaluable.

Recall the expression used to approximate the variance of improved estimates; see Expression \ref{varest2}. In the simulation studies presented in this paper, negative estimates of the variance of improved estimates are evaluated for some estimators. Future work is required to address this issue. Three potential approaches are to: 1) Opt to use a conservative approach towards estimating the variance of the preliminary estimator so that $E[\hat{\text{var}}(\hat{\gamma}_0)|d_r]$ is likely to be larger than $\text{var}(\hat{\gamma}_0|d_r)$, and hence the estimate is more likely to be positive; 2) Base a jackknife-type variance estimator solely on a series of Rao-Blackwellized estimates that correspond to a series of subsets of elements removed from the original samples. However, one drawback with this approach is that it will require more computational resources relative to current approaches since one will be required to evaluate approximations for each of the improved estimators; 3) For Bayesian- and bootstrap-based estimates, where cutoff points of the posterior and resampling distributions can be used to form alternative probabiliy/confidence intervals, base such intervals on the these values as they appear to behave well in the simulation studies.

Developments on extending the theoretical results presented in this paper to more elaborate mark-recapture models would be useful. Such models may consist of those based on a two- or three-pair combination of heterogeneity, behavioural, and time effects, as well as the assumption that the population is open. Another valuable contribution would be to extend the strategy to work over mark-recapture models that allow for interaction effects between the sampling occasions \citep{Baillargeon2007, Chan2019}. Future work is being prepared for such models.

%%%%%%  bibliography

%\newpage
\bibliographystyle{biom}
%\addcontentsline{Bibliography}
\bibliography{MasterReferences}

\clearpage

\appendix

\textbf{\Large Supplementary Materials}

The supplementary materials accompany the main paper ``Rao-Blackwellization to Give Improved Estimates in Multi-List Studies". The stratified mark-recapture model and corresponding inference procedure is detailed. Results from applying the Rao-Blackwell inference strategy to empirical data sets are also presented.

\section{Stratified Setup}

Suppose there are $G$ strata the population is partitioned into; typically, such partitions would be based on covariate classes like a crossing of age category with gender. Further suppose unit $i$ belongs to some stratum $j$ that can be observed upon capturing unit $i$, where  $j=1,2,...,G$. Define $p_{ik}$ to be the capture probability of unit $i$ on sampling occasion $k$. Under the stratified setup for the this model, $p_{ik}=p_j$ for all $i=1,2,...,N$ and $k=1,2,...,K$, where $p_j$ is the capture probability corresponding with units from stratum $j$.

\subsection{Sufficiency Result}

When considering unit labels, define the original data to be $d_0=\{s_k:k=1,...,K\}$ and the reduced data to be $d_R=\{s,\underline{C}'\}$, where $s=\cup_{k=1}^K s_k$, $\underline{C}'=(C_1',C_2',...,C_G')$, $C_j'=\sum\limits_{k=1}^K C_{j,k}$, $C_{j,k}$ is the number of units captured from stratum $j$ on sampling occasion $k$, and $C_{i,j,k}=1$ if unit $i$ from stratum $j$ is captured on sampling occasion $k$ and 0 otherwise. Define $n=|s|$. Define $s_{(j)}$ to be the set of units captured from stratum $j$ at least once in the study, and $n_{(j)}=|s_{(j)}|$.

\textbf{Theorem:} The reduced data $D_R$ is sufficient for $(\underline{N},\underline{p})$ where $\underline{N}=(N_1,N_2,...,N_G)$ (that is, the corresponding sizes of each stratum) and $\underline{p}=(p_1,p_2,...,p_G)$.

\textbf{Proof:} For any $(d_0,\underline{p})$, the labels-based likelihood is
\begin{align}
\notag
&P_{\underline{N}}(D_0=d_0)=P(s_1,s_2,...,s_K)I[s\subseteq U]\\\notag
&=\prod\limits_{k=1}^K\bigg[\prod\limits_{j=1}^Gp_j^{\sum\limits_{i \epsilon s_{(j)}}C_{i,j,k}}
(1-p_j)^{\sum\limits_{i\epsilon s_{(j)}} (1-C_{i,j,k})}  (1-p_j)^{(N_j-n_{(j)})}   \bigg]I[s\subseteq \{1,2,...,N\}]\\\notag
&=\prod\limits_{j=1}^G\bigg[p_j^{\sum\limits_{k=1}^K\sum\limits_{i \epsilon s_{(j)}}C_{i,j,k}}
(1-p_j)^{\sum\limits_{k=1}^K\sum\limits_{i\epsilon s_{(j)}} (1-C_{i,j,k})}  (1-p_j)^{(KN_j-Kn_{(j)})}   \bigg]I[s\subseteq \{1,2,...,N\}]\\\notag
&=\prod\limits_{j=1}^G\bigg[p_j^{\sum\limits_{k=1}^KC_{j,k}}
(1-p_j)^{\sum\limits_{k=1}^K (n_{(j)}-C_{j,k})}  (1-p_j)^{(KN_j-Kn_{(j)})}   \bigg]I[s\subseteq \{1,2,...,N\}]\\\notag
&=\prod\limits_{j=1}^G\bigg[p_j^{C_{j}'}(1-p_j)^{(KN_j-C_j')}  \bigg]I[s\subseteq \{1,2,...,N\}]\\\notag
&=g(\underline{N},\underline{p},d_R)h(d_0)\\
\label{sufficiency_Mh}
\end{align}
where $h(d_0)=1$. Therefore, by the Neyman-Factorization Theorem $D_R$ is sufficient for $(\underline{N},\underline{p})$. $\Box$

Table \ref{strat_reordering} depicts a sample selected under a stratified setup (where A corresponds with unit 1, B with unit 2, and C with unit 3), where the probability of observing the original data is $p_1^4(1-p_1)^{(3N_1-4)}\times p_2^2(1-p_2)^{(3N_2-2)}$, and a sample reordering that is consistent with the reduced data.

\begin{table}[H]
\centering
\caption{Left: Original sample selected under a stratified setup. Right: Sample reordering consistent with the reduced data. Letters refer to units and numbers to stratum membership.}
\begin{tabular}{l||lll||llllll}\hline
Unit                        &Sample 1 &Sample 2   &Sample 3       &Sample 1 &Sample 2   &Sample 3 \\\hline
$A_1$                         &1        &1          &0              &1        &1          &1      \\
$B_1$                         &0        &1          &1              &1        &0          &0      \\
$C_2$                         &1        &1          &0              &1        &0          &1     \\
\hline
\label{strat_reordering}
\end{tabular}
\end{table}

\subsection{Resampling Procedure}

A Metropolis-Hastings MCMC chain \citep{Hastings1970} is used to approximate the improved estimators with the following proposal distribution. Essentially, the proposal distribution based on the null model is adopted and applied to each stratum as follows.

Step 1: For each of the $n_{(j)}$ individuals in $s_{(j)}$, assign them a capture to a randomly chosen sample and tabulate the capture history matrix.

Step 2: Distribute the remaining $\sum\limits_{k=1}^Kn_{(j,k)} - n_{(j)}$ captures at random to the zero entries in the capture history matrix corresponding to the entries from stratum $j$, where $n_{(j,k)}$ is the number of units from stratum $j$ captured on sampling occasion $k$.

In the first step and for each captured unit $i$, there are $C_i$ possible ways of first assigning this unit to one of the $C_i$ samples in which they are captured. Therefore, the total possible number of ways of selecting the proposed sample reordering is $\prod\limits_{i\ \epsilon\ s} C_i$. For the accept/reject portion of the chain, the sample reordering is accepted with probability $\text{min}\bigg\{\frac{q(d_0^{[r^*]})}{q(d_0^{[r]})},1\bigg\}=\text{min}\bigg\{\frac{\prod\limits_{i\ \epsilon\ s} C_i^*}{\prod\limits_{i\ \epsilon\ s} C_i}, 1\bigg\}$ where $d_0^{[r^*]}$ and $d_0^{[r]}$ respectively correspond to the most recently accepted sample reordering and proposed sample reordering.

\subsection{Simulation Study}

The population size is set to $N=500$. The total number of sampling occasions is five. The population is partitioned into three strata/covariate classes of size 200, 200, and 100.  The respective capture probabilities for individuals within the covariate classes are set to 0.25, 0.275, and 0.30. The following estimator, which is already a function of the sufficient statistic, is used in the simulation study: the Huggins estimator \citep{Huggins1989, Hwang2005} based on the assumption of capture effects by covariate class. The following estimators, which are not functions of the sufficient statistic and can therefore benefit from the Rao-Blackwellization procedure detailed in this section, are also used in the simulation study: the Pledger mixture model maximum likelihood estimator based on two mixtures \citep{Pledger2000} and the assumption of capture effects by covariate class, and the Chao-Tsay and Chao-Bunge sample coverage approach estimators. Table \ref{Sim_Stratified} presents results from the simulation study. Improved estimates are based on 1000 resamples. The acceptance rate of the MCMC chain is 22.85\%.

\begin{longtable}{l*{10}{r}r}
\caption{Mean, variance, mean-squared error, coverage rates, average length of confidence intervals, coverage rates of alternative confidence intervals, average length of alternative confidence intervals, and percentage of variance estimates which are negative, for preliminary and improved estimators. Top: Estimators which are already functions of the sufficient statistic. Middle: Preliminary estimators. Bottom: Improved estimators.}
\endfirsthead
\multicolumn{10}{l}
{{Table \ref{Sim_Stratified} continued from previous page}} \\
  \hline
Estimator       &Mean           &Var.           &MSE                &CR      &Length   &Alt. CR       &Alt. Length         &Neg. Est.             \\\hline
\endhead
Estimator       &Mean           &Var.           &MSE                &CR      &Length   &Alt. CR       &Alt. Length         &Neg. Est.            \\\hline
Huggins         &503            &279            &284                &0.949   &65       &0.994         &111                 &NA              \\\hline

Pledger         &497            &332            &343                &0.927   &73       &0.950         &74                  &NA              \\
SC CT           &501            &501            &502                &0.943   &87       &NA            &NA                  &NA              \\
SC CB           &501            &572            &573                &0.948   &93       &0.946         &95                  &NA              \\\hline

Pledger RB      &496            &288            &300                &0.943   &72       &0.964        &74                   &0\%                 \\
SC CT RB        &501            &303            &304                &0.945   &68       &NA           &NA                   &0.2\%                   \\
SC CB RB        &501            &309            &310                &0.943   &69       &0.987        &95                   &0.68\%               \\\hline
\label{Sim_Stratified}
\end{longtable}
\vspace*{-\baselineskip}

\section{Empirical Applications}

The new inference strategy is applied to several empirical data sets. Results based on each data set are presented in the following subsections.

\subsection{$M_0$ and $M_h$ Model Application: Diabetes Data Set}

The new strategy is applied to a diabetes data set which is based on four administrative records from a community in Italy \citep{Bruno1994}. Capture histories can be found in \cite{Chao2001}.

The data set has been analysed by \cite{IWGDMF-1, IWGDMF-2} with the use of a mark-recapture model that assumes heterogeneity is present in the captures. \cite{Chao2001} analyse this data set and with the sample coverage approach their proposed estimate is 2,609 with variance estimate based on bootstrap replications of 6,561 and corresponding confidence interval of (2,472; 2,792).

Tables \ref{Empirical_M0} and \ref{Empirical_Mh} provide estimates respectively based on the $M_0$ and $M_h$ model. For the $M_0$ estimators, 250,000 resamples are used to approximate the improved estimators and the acceptance rate is 0.01\% For the $M_h$ model, 5000 resamples are used to approximate the improved estimators. Most of the resulting confidence intervals corresponding to the Rao-Blackwellized estimators capture the population size estimate suggested by \cite{Chao2001}, and most provide substantial increases in the estimated precision.

\begin{longtable}{l*{10}{r}r}
\caption{Population size estimates corresponding to diabetes data set based on new strategy, $M_0$ model assumption. Point estimate, variance estimate, confidence intervals and alternative confidence intervals for preliminary and improved estimators, and if a negative estimate was initially obtained for the variance estimate of the improved estimator. Top: Estimators which are already functions of the sufficient statistic. Middle: Preliminary estimators. Bottom: Improved estimators.}
\endfirsthead
\multicolumn{10}{l}
{{Table \ref{Empirical_M0} continued from previous page}} \\
  \hline
Estimator           &Estimate     &Variance Estimate    &CI                 &Alt CI.                &Neg. Est.           \\\hline
\endhead
Estimator           &Estimate     &Variance Estimate    &CI                 &Alt CI.                &Neg. Est.          \\\hline
MLE                 &2,525        &1,054                &(2,462; 2,589)     &(2,466; 2,593)         &NA                  \\
MLE LL              &2,526        &1,055                &(2,462; 2,590)     &NA                     &NA                 \\
Bayes               &2,525        &1,048                &(2,462; 2,589)     &(2,464; 2,591)         &NA                 \\\hline

LP                  &2,351        &3,345                &(2,238; 2,464)     &NA                     &NA                  \\
JK                  &3,218        &5,850                &(3,068; 3,368)     &NA                     &NA         \\
SC CT               &2,458        &2,467                &(2,361; 2,555)     &(2,372; 2,568)         &NA          \\\hline

LP RB               &2,515        &8,314                &(2,336; 2,693)     &NA                     &Yes            \\
JK RB               &3,231        &4,397                &(3,101; 3,361)     &NA                     &No              \\
SC CT RB            &2,512        &1,107                &(2,447; 2,578)     &(2,428; 2,616)         &No          \\\hline
\label{Empirical_M0}
\end{longtable}
\vspace*{-\baselineskip}

\begin{longtable}{l*{10}{r}r}
\caption{Population size estimates corresponding to diabetes data set based on new strategy, $M_h$ model assumption. Point estimate, variance estimate, confidence intervals and alternative confidence intervals for preliminary and improved estimators, and if a negative estimate was initially obtained for the variance estimate of the improved estimator. Top: Estimators which are already functions of the sufficient statistic. Middle: Preliminary estimators. Bottom: Improved estimators.}
\endfirsthead
\multicolumn{10}{l}
{{Table \ref{Empirical_Mh} continued from previous page}} \\
  \hline
Estimator           &Estimate     &Variance Estimate    &CI                 &Alt CI.                &Neg. Est.           \\\hline
\endhead
Estimator           &Estimate     &Variance Estimate    &CI                 &Alt CI.                &Neg. Est.          \\\hline
Pledger             &2,559        &1,264                &(2,489; 2,629)     &(2,494; 2,634)         &NA                  \\
Chao                &2,513        &1,497                &(2,437; 2,589)     &NA                     &NA                 \\
Poisson             &2,591        &2,398                &(2,495; 2,687)     &NA                     &NA                 \\
JK                  &3,218        &5,850                &(3,068; 3,368)     &NA                     &NA                 \\
SC CB               &2,546        &1,891                &(2,468; 2,639)     &NA                     &NA                  \\\hline

Bayes               &2,632        &5,293                &(2,489; 2,774)     &(2,514; 2,778)         &NA          \\
SC CT               &2,458        &2,493                &(2,360; 2,556)     &(2,372; 2,569)         &NA          \\\hline

Bayes RB            &2,641        &2,663                &(2,539; 2,742)     &(2,540; 2,773)         &No            \\
SC CT RB            &2,556        &2,536                &(2,457; 2,654)     &(2,467; 2,665)         &No        \\\hline
\label{Empirical_Mh}
\end{longtable}
\vspace*{-\baselineskip}

\subsection{$M_t$ Model Application: Hare Data Set}

The new strategy is applied to a population of snowhshoe hare data with six capture occasions \citep{Cormack1989}. Capture histories can be found in the `Rcapture' package \citep{Baillargeon2007, Rivest2012}.

\cite{Baillargeon2007} analyse this data set and find that a large degree of heterogeneity is introduced by two hares which are captured for all six sampling occasions. Consequently, they suggest removing them from the data set for estimation purposes. Based on goodness-of-fit criteria, they suggest using the $M_t$ model. The resulting estimate of the population size is 76.78 with variance estimate 15.30 and confidence interval based on the profile likelihood method of (70.09; 85.41).

Table \ref{Empirical output Hare} provides the estimates based on the $M_t$ model. A total of 2000 resamples are used to approximate the improved estimators and the acceptance rate is 83.65\%. Most of the resulting confidence intervals corresponding to Rao-Blackwellized estimators capture the population size estimate suggested by \cite{Baillargeon2007}, and each provides a substantial increase in the estimated precision.

\begin{longtable}{l*{10}{l}{r}r}
\caption{Population size estimates corresponding to hare data set based on new strategy, $M_t$ model assumption. Point estimate, variance estimate, confidence intervals and alternative confidence intervals for preliminary and improved estimators, and if a negative estimate was initially obtained for the variance estimate of the improved estimator. Top: Estimators which are already functions of the sufficient statistic. Middle: Preliminary estimators. Bottom: Improved estimators.}
\endfirsthead
\multicolumn{10}{l}
{{Table \ref{Empirical output Hare} continued from previous page}} \\
  \hline
Estimator           &Estimate     &Variance Estimate    &CI                 &Alt CI.                &Neg. Est.\\\hline
\endhead
Estimator           &Estimate     &Variance Estimate    &CI                 &Alt. CI                &Neg. Est.    \\\hline
MLE                 &74.05        &14.78                &(66.51; 81.58)     &(69.31; 85.57)         &NA            \\
MLE LL              &75.89        &17.70                &(67.65; 84.14)     &NA                     &NA            \\
Bayes               &74.85        &15.66                &(67.09; 82.60)     &(68.00; 84.00)         &NA            \\\hline

LP                  &134.00       &3,240.00             &(22.43; 245.57)    &NA                     &NA            \\
JK                  &91.00        &50.00                &(77.00; 105.00)    &NA                     &NA            \\
SC CB               &78.00        &40.38                &(70.00; 98.00)     &NA                     &NA            \\
SC CT               &80.06        &54.10                &(65.64; 94.47)     &(71.36; 102.87)        &NA            \\\hline

LP RB               &74.50        &116.85               &(53.32; 95.69)     &NA                     &No            \\
JK RB               &88.69        &29.93                &(78.00; 99.41)     &NA                     &No             \\
SC CB RB            &75.10        &15.21                &(67.45; 82.74)     &NA                     &No             \\
SC CT RB            &75.32        &17.31                &(67.17; 83.48)     &(69.17; 94.64)         &No             \\\hline
\label{Empirical output Hare}
\end{longtable}

\subsection{$M_t$ Model Application: HIV Data Set}

The new strategy is applied to an epidemiological four list/sample study based on an HIV population in Rome, Italy \citep{Abeni1994}. Capture histories can be found in the `Rcapture' package \citep{Baillargeon2007, Rivest2012}.

\cite{Baillargeon2007} analyse this data set and based on goodness-of-fit criteria suggest using the $M_t$ model with interaction terms between the first two lists. The resulting estimate of the population size is 12,319, with variance estimate 1,413,060 and confidence interval based on the profile likelihood method of (10,287; 14,978).

Table \ref{Empirical output HIV} provides the estimates based on the $M_t$ model. A total of 15,000 resamples are used to approximate the improved estimators and the acceptance rate is 83.65\%. Most of the resulting confidence intervals corresponding to Rao-Blackwellized estimators capture the population size estimate suggested by \cite{Baillargeon2007}, and most provide substantial increases in the estimated precision.

\begin{longtable}{l*{10}{l}{r}r}
\caption{Population size estimates corresponding to hare data set based on new strategy, stratified model assumption. Point estimate, variance estimate, confidence intervals and alternative confidence intervals for preliminary and improved estimators, and if a negative estimate was initially obtained for the variance estimate of the improved estimator. Top: Estimators which are already functions of the sufficient statistic. Middle: Preliminary estimators. Bottom: Improved estimators.}
\endfirsthead
\multicolumn{10}{l}
{{Table \ref{Empirical output HIV} continued from previous page}} \\
  \hline
Estimator           &Estimate     &Variance Estimate    &CI                 &Alt CI.                &Neg. Est.\\\hline
\endhead
Estimator           &Estimate     &Variance Estimate    &CI                 &Alt. CI                &Neg. Est.    \\\hline
MLE                 &11,117       &817,396              &(9,345; 12,889)    &(9,508; 13,066)        &NA         \\
MLE LL              &11,069       &803,667              &(9,312; 12,826)    &NA                     &NA         \\
Bayes               &11,187       &805,991              &(9,427; 12,946)    &(9,549; 13,081)        &NA         \\\hline

LP                  &7,754        &1,331,148            &(5,492; 10,015)    &NA                     &NA         \\
JK                  &5,329        &10,644               &(5,127; 5,531)     &NA                     &NA         \\
SC CB               &19,430       &105,352,673          &(7,947; 52,705)    &NA                     &NA         \\
SC CT               &12,464       &1,668,197            &(9,932; 14,995)    &(10,220; 15,312)       &NA         \\\hline

LP RB               &12,727       &1,348,014             &(10,451; 15,002)  &NA                     &No         \\
JK RB               &5,184        &10,065                &(4,988; 5,831)    &NA                     &Yes         \\
SC CB RB            &11,056       &5,286,812             &(6,550; 15,563)   &NA                     &No          \\
SC CT RB            &11,042       &812,532               &(9,275; 12,808)   &(9,153; 13,422)        &No          \\\hline
\label{Empirical output HIV}
\end{longtable}

\subsection{Stratified Model Application: Dipper Data Set}

The new strategy is applied to a seven sample study based on European dippers from France and the capture histories can be found in the `RMark' package \citep{RMark}. For each captured unit, the gender is recorded and this serves as a stratification variable.

Table \ref{Empirical dipper} provides the estimates based on the stratified model. A total of 2,500 resamples are used to approximate the improved estimators and the acceptance rate is 18.72\%. Significant improvements in the (estimated) standard error are found with the new inference strategy.

\begin{longtable}{l*{10}{l}{r}r}
\caption{Population size estimates corresponding to dipper data set based on new strategy. Point estimate, variance estimate, mean-squared error, confidence intervals and alternative confidence intervals for preliminary and improved estimators, and if a negative estimate was initially obtained for the variance estimate of the improved estimator. Top: Estimators which are already functions of the sufficient statistic. Middle: Preliminary estimators. Bottom: Improved estimators.}
\endfirsthead
\multicolumn{10}{l}
{{Table \ref{Empirical dipper} continued from previous page}} \\
  \hline
Estimator           &Estimate     &Variance Estimate    &CI                 &Alt CI.                &Neg. Est.\\\hline
\endhead
Estimator           &Estimate     &Variance Estimate    &CI                 &Alt. CI                &Neg. Est.    \\\hline
Huggins             &373          &190                  &(346; 400)         &(344; 421)             &NA             \\\hline

Pledger             &447          &1,917                &(362; 533)         &(383; 560)             &NA              \\
SC CB               &527          &5,005                &(424; 711)         &NA                     &NA               \\
SC CT               &553          &2,155                &(462; 644)         &(477; 661)             &NA              \\\hline

Pledger RB          &376          &521                  &(331; 421)         &(346; 429)             &No               \\
SC CB RB            &377          &253                  &(345; 408)         &NA                     &No                 \\
SC CT RB            &377          &236                  &(347; 407)         &(345; 428)             &No               \\\hline
\label{Empirical dipper}
\end{longtable}

\end{document}